\documentstyle[epsfig]{mn}

\begin{document}
\def\ltsima{$\; \buildrel < \over \sim \;$}
\def\simlt{\lower.5ex\hbox{\ltsima}}
\def\gtsima{$\; \buildrel > \over \sim \;$}
\def\simgt{\lower.5ex\hbox{\gtsima}}

\title[Changing look: from Compton--thick to Compton--thin]
{Changing look: from Compton--thick to Compton--thin, or the
re--birth of fossil AGN}

\author[Giorgio Matt, Matteo Guainazzi and Roberto Maiolino]
{Giorgio Matt$^1$, Matteo Guainazzi$^2$ and Roberto Maiolino$^{1,3}$\\ ~ \\
$^1$Dipartimento di Fisica, Universit\'a degli Studi ``Roma 
Tre'', Via della Vasca Navale 84, I--00146 Roma, Italy \\
$^2$XMM Science Operation Center, RSSD--ESA, 
VILSPA, Apartado 50727, E--28080 Madrid, Spain\\
$^3$INAF--Osservatorio Astrofisico di Arcetri, Largo Fermi 5, 50125 
Firenze, Italy\\
}

\maketitle
\begin{abstract}
We discuss the properties of a small sample of Seyfert 2 galaxies
whose X--ray spectrum changed appearance on time scales of years,
becoming reflection--dominated from Compton--thin, or viceversa.
A reflection--dominated spectrum is usually taken as evidence of   
Compton--thick absorption, but we instead argue that such a spectrum
is due to a temporary switching--off
of the nuclear radiation. The observations discussed here may help
explaining mismatches between optical and X--ray classifications, and
provide new strong and direct evidence
of the presence of more than one cold circumnuclear
region in Seyfert 2 galaxies.
\end{abstract}


\section{Introduction}

\begin{figure}[t]
\epsfig{file=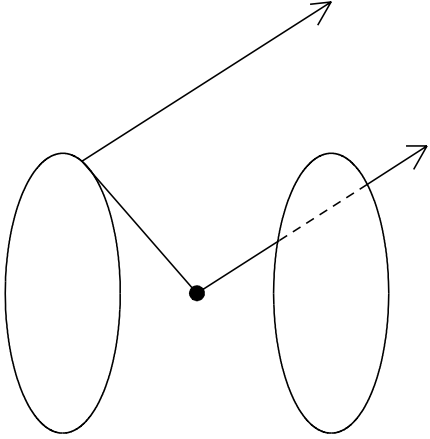,height=90mm,width=90mm}
\caption{Sketch of the geometry adopted for the calculations presented
in Figs.~2-4. The opening angle of the torus is  30$^{\circ}$.}
\label{torus}
\end{figure}

\begin{figure}[t]
\epsfig{file=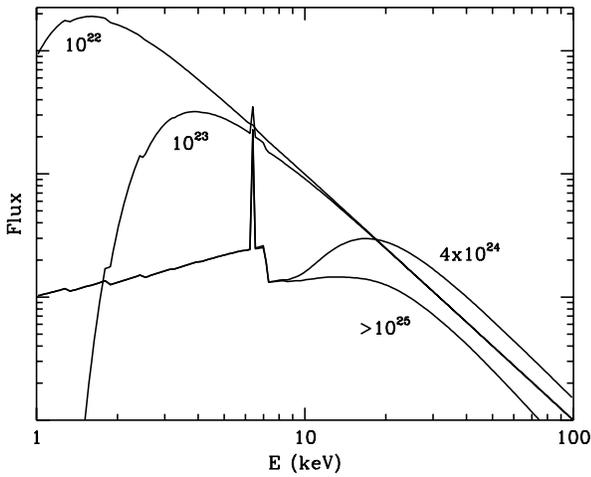, height=9cm,width=9cm}
\caption{ The X--ray spectrum of an obscured AGN is
shown, for different column densities of the absorber. The absorbing
matter is assumed to form
a geometrically thick torus, according to Unification
models. } 
\label{cthall}
\end{figure}

\begin{figure}[t]
\epsfig{file=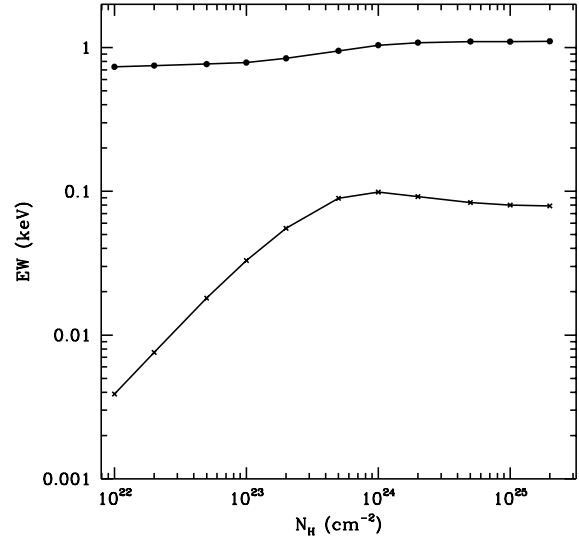,height=90mm,width=90mm}
\caption{For the same geometry and illuminating spectrum as
in Fig.~\ref{refl},
the EW of the iron line against the pure reflection component (upper data)
and against the total continuum (lower data). The torus is assumed to be 
face--on.}
\label{reflew}
\end{figure}

It is well known (see e.g. Matt 2002 for a review, and references therein) 
that most AGN are `obscured' in X--rays.  Their observed spectrum
depends on the (hydrogen equivalent) column density, $N_{\rm H}$, 
of the absorber. If the column density exceeds the value,
$\sigma_T^{-1}$=1.5$\times$10$^{-24}$
cm$^{-2}$, for which the Compton scattering optical depth becomes
equal to 1, the sources are called `Compton--thick'. If the column
density is smaller than $\sigma_T^{-1}$ but still in excess of the
Galactic one, the source is called `Compton--thin'. Assuming the
simple geometry depicted in Fig.~\ref{torus} (in which the absorbing
matter is assumed to form a geometrically thick torus, according
to the popular Unification models, see Antonucci 1993), 
the expected spectrum is shown in Fig.~\ref{cthall}. 
In the Compton--thin case,
the nuclear spectrum (assumed for simplicity to
be a power law with photon spectral index 2)
can be directly observed above a few keV,
and a fluorescent iron line (with EW$\sim$10 eV for N$_H$=10$^{22}$ 
cm$^{-2}$ and
$\sim$100 eV for N$_H$=10$^{23}$ cm$^{-2}$, see Fig.~\ref{reflew}),
is produced.
In the moderately Compton--thick case (N$_H$=$4\times$10$^{24}$ cm$^{-2}$)
the spectrum below 10 keV is dominated by the reflection continuum 
(plus a prominent iron line with 
EW$\sim$1 keV: Ghisellini et al. 1994; Krolik et al. 1994; Matt et al. 1996)
produced by the visible part of the inner wall of the torus itself 
(see Fig.~\ref{torus}), while the nuclear radiation can
be directly visible in transmission at higher energies. This is, for instance,
the case of two out of the three closest AGN,
the Circinus Galaxy (Matt et al. 1999) and NGC~4945 (Iwasawa et al. 1993;
Guainazzi et al. 2000). For column densities
exceeding  N$_H$=10$^{25}$ cm$^{-2}$, no nuclear radiation is transmitted,
and only the reflection component is visible (see for instance NGC~1068,
Matt et al. 1997). It is worth noting that
reflection from highly ionized matter can also be present (cf. Matt et al.
2000), and that the ionization state of the ``cold" reflector may include also
mildly ionized components (Bianchi et al. 2001). However, for simplicity
we will discuss cold reflectors only.

\begin{figure}[t]
\epsfig{file=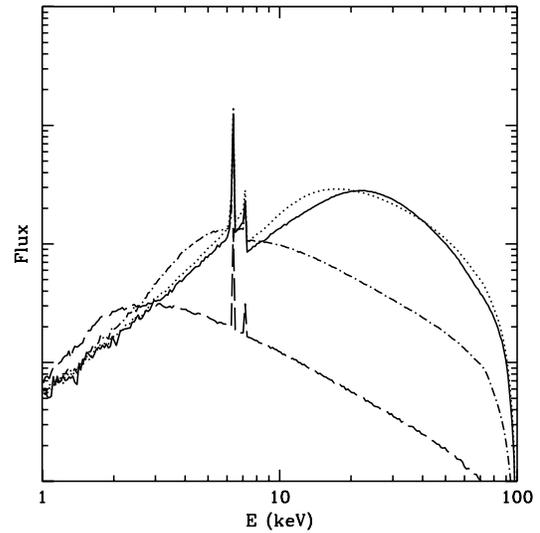,height=90mm,width=90mm}
\caption{ The reflection spectrum from the torus
for different column densities:
2$\times10^{22}$ cm$^{-2}$ (dashed line), 2$\times10^{23}$
cm$^{-2}$ (dotted--dashed line), 2$\times10^{24}$ cm$^{-2}$ (dotted line),
2$\times10^{25}$ cm$^{-2}$ (solid line).
The illuminating spectrum is a power law with photon index 2 and
exponential cut--off at 100 keV.  The torus is assumed to be 
face--on.}
\label{refl}
\end{figure}

Also the spectrum reflected by the torus depends on its column
density. The iron line EW (calculated with
respect to both the reflection continuum and the 
total one), is shown in Fig.~\ref{reflew} for a torus  
seen face--on as a function of the $N_{\rm H}$ in the equatorial plane. 
In Fig.~\ref{refl} the reflection spectrum is shown. For Compton--thick
material, the spectrum is well--known (e.g. George \& Fabian 1991;
Matt et al. 1991) while for Compton--thin matter it is very different in 
shape and, overall, less prominent. (Both figures are based on Monte Carlo
simulations; see Ghisellini et al. 1994 for details on the simulation
code.)

A reflection--dominated spectrum is usually assumed as evidence
of Compton--thick absorption, expecially when using
instruments working up to $\sim$10 keV, and therefore
unable to detect the primary emission through moderately thick absorbers.
However, such a spectrum may occur not only when the nucleus is absorbed,
but also when its emission decreases down to invisibility 
(in this case we will speak of a `switched--off' source), 
provided of course that the reflecting
material is distant enough from the nucleus to act as an `echo' for
a while. The classical example of a switched--off source is NGC~4051,
observed by BeppoSAX during a prolonged low state (Uttley et al. 1999)
and found to be reflection--dominated (Guainazzi et al. 1998), i.e.
with a very prominent iron line and the hard (and curved) continuum
spectrum typical of a Compton--thick reflection component. NGC~4051
is a well--known Seyfert 1 galaxy but, based on the BeppoSAX spectrum
alone, it would have been classified as a Compton--thick absorbed source.
The explanation in terms of a decrease of the nuclear flux
instead of a temporary Compton--thick absorption is further confirmed
by recent Chandra observations (Fruscione et al. 2002), which showed a
residual short--term and low amplitude variability during another low state
of the source; clearly, in this case the nuclear flux was much
fainter then usual, but not completely disappeared as during the 
BeppoSAX observation.
We note, incidentally, that a temporary switching--off of the nucleus
may explain at least some of the type 1 sources with a large hardness
ratio discovered in X--ray surveys, see e.g. Fiore et al. (2001) and
Della Ceca et al. (2001).

Recently, the same kind of variability was discovered
in a few Seyfert 2 galaxies,
which changed from reflection--dominated (and therefore candidate
Compton--thick absorbed sources) to Compton--thin, or viceversa.
The clearest cases so far are UGC~4203 (Guainazzi et al. 2002), 
NGC~6300 (Guainazzi 2002), NGC~1365 (Risaliti et al. 2000)
and NGC~2992 (Gilli et al. 2000). 
For at least some of these sources, a change in the column density
of the absorber, rather than a switching--off of the source, cannot
be completely ruled out, and indeed Risaliti et al. (2002) claimed
that variations in the absorbing column density are  common 
in Seyfert 2 galaxies.
However, the changing absorbers in the Risaliti et al. (2002) sample are all
Compton--thin, and in most cases the variations are too small to
rule out the possibility that they are an artifact due to comparing
spectra obtained with different instruments. Moreover, this solution is
clearly untenable for NGC~2992 (Gilli et al. 2000; see Sec.~2.4), 
which has been well monitored over the years, 
showing a gradual change of the nuclear flux and a costant
absorber. We find difficult
to imagine a situation in which a Compton--thick absorber on the pc--scale
(as suggested by the lack of variability of the reflection components)
and with a large covering
factor (to allow for the rather large reflection components) can vary
so dramatically on time--scales of years. Therefore, in the following we
will assume that the observed variations are due to the switching--off of the
nucleus. After reviewing the current
observation status of this field (Sect.~2), we will discuss some possible
implications (Sect.~3).

\section{Seyfert 2 galaxies which changed look}

\subsection{UGC~4203}

UGC~4203 (a.k.a. Mkn~1210) has been
recently observed by XMM--Newton (Guainazzi et al. 2002), unveiling 
a X--ray bright nucleus,
absorbed by $N_H \simeq 2 \times 10^{23}$~cm$^{-2}$. However,
in an ASCA observation performed about five and half years
earlier (Awaki et al. 2000), the prominent iron line ($EW \simeq 1$~keV) and
the factor of 5 lower 2--10~keV flux indicated 
a reflection--dominated spectrum (Fig.~\ref{polittico}), with the nuclear
emission too faint to be visible.

The limited bandpass of ASCA, along with the low flux of the source,
does not permit to distinguish bewteen different column densities of
the reflecting matter, provided that  it 
exceeds about $10^{23}$~cm$^{-2}$ (see Fig.~\ref{refl}). 
It is therefore
possible that in this case the absorbing and reflecting materials
are one and the same.

\subsection{NGC~6300}

NGC~6300, discovered serendipitously by {\it Ginga}
(Awaki et al. 1991), was observed in a Compton--thick reflection-dominated
state by RXTE on February 1997 (Leighly et al. 1999).
Two and half years later a remarkably strong Seyfert nucleus
(2--10~keV flux $\sim 1.3 \times 10^{-11}$~erg~cm$^{-2}$~s$^{-1}$)
seen through a column density with $N_H \simeq 2 \times 10^{23}$~cm$^{-2}$
(Fig.~\ref{polittico}), was discovered in a BeppoSAX observation.  
An XMM--Newton observation performed early in 2001
caught the source still in the high flux, Compton--thin state (Maddox et al.
2002).

As the RXTE  bandpass extends up to 20 keV, for this source it is possible
to distinguish between 
Compton--thin and Compton--thick reflection (Fig.~\ref{refl}). 
The detection in the PCA highest
energy band is too strong to be explained as pure reflection by matter
with $N_H\simeq 2 \times 10^{23}$~cm$^{-2}$.
In this case, therefore, the (thick) reflector must be different
from the (thin) absorber.

\subsection{NGC~1365}

A BeppoSAX observation on August 1997 detected in this 
source a bright Seyfert nucleus, seen through
a Compton--thin ($N_H \simeq 4 \times 10^{23}$~cm$^{-2}$;
Risaliti et al. 2000) absorber. On the contrary, an
ASCA observation, performed three years earlier,
detected a very flat 
X-ray continuum ($\Gamma \simeq 0.8$) and
a 2.1~keV K$\alpha$ iron line, both
indicating a reflection--dominated state (Fig.~\ref{polittico}). 
Due to the limited bandwidth of ASCA, and similarly to UGC~4203, 
the possibility that the reflector
is simply the inner wall of the absorber cannot be ruled out.

\subsection{NGC~2992}

\begin{figure*}
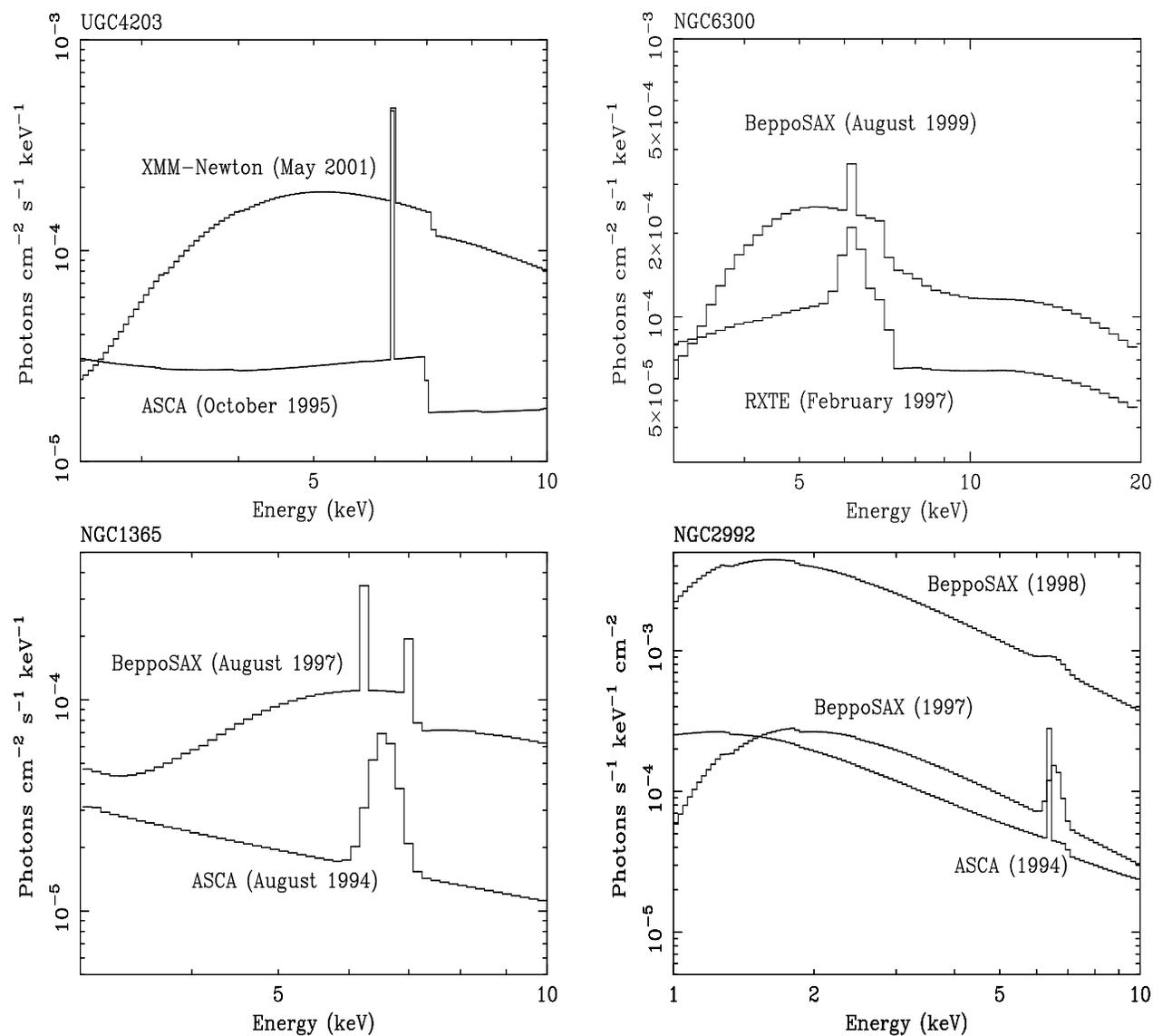

\hbox{
\includegraphics*[width=7.5cm,height=8.0cm,angle=-90]{asca_vs_xmm_model.ps}
\hspace{0.5cm}
\includegraphics*[width=7.5cm,height=8.0cm,angle=-90]{bsax_vs_rxte_model.ps}
}
\hbox{
\includegraphics*[width=7.5cm,height=8.0cm,angle=-90]{ngc1365.ps}
\hspace{0.5cm}
\includegraphics*[width=7.5cm,height=8.0cm,angle=-90]{ngc2992.ps}
}
\caption{Upper-left panel: The ASCA and XMM--Newton  best-fit models
for UGC~4203 (Guainazzi et al. 2002). Upper-right panel: 
The BeppoSAX and RXTE best-fit models
for NGC~6300 (after Leighly et al. 1999 and Guainazzi 2002). 
Lower-left panel: ASCA and BeppoSAX best-fit models 
for NGC~1365 (after 
Iyomoto et al. 1997 and Risaliti et al. 2000, respectively). Note
that, in agreement with these papers,
the line feature is represented as a single broad Gaussian profile at $E =
6.57$~keV (ASCA) or as the blending of two narrow features (BeppoSAX) at
energies $E = 6.257 \pm 0.09$~keV and 6.95~keV, respectively.
Lower-right panel: ASCA (1994) and BeppoSAX (1997,
1998) best fit models of NGC~2992 
(after Waever et al. 1996 and Gilli et al. 2000).
See text for mode details on each source. }
\label{polittico}
\end{figure*}

The brightest and best studied source in our little sample is NGC~2992,
a Seyfert 1.9 galaxy with an X--ray absorbing column density 
N$_{H}\sim9\times10^{21}$ cm$^{-2}$.
The X--ray flux of NGC~2992 steadily declined since 1978, 
when it was observed by HEAO--1 (Mushotzky 1982) at a (unabsorbed)
flux level of about 8$\times$10$^{-11}$ erg cm$^{-2}$ s$^{-1}$,
until 1994, when it was observed by ASCA
(Weaver et al. 1996) at a flux level more than one order of magnitude fainter.
Then the source underwent a rapid recovery: in 1997 it was observed by
BeppoSAX
at a flux level somewhat higher than in 1994, while in 1998 it fully recovered
its 1978 brightness (Gilli et al. 2000; see Fig.~\ref{polittico}).

In this case the
comparison between the Compton-thin and the ASCA almost
reflection-dominated states
clearly rules out the possibility that the reflector is the inner wall of the
absorber. The 4-10 to 2-4 keV flux ratio during the ASCA observation is $0.35
\pm 0.02$ (corresponding to $\Gamma \simeq 1.15$), which is largely
inconsistent with the theoretical value expected from a reflection dominated
spectrum by a column density $\sim 10^{22}$~cm$^{-2}$ (0.53). It is
worth noting that in both BeppoSAX observations the power law spectral
index was typical for AGN
($\Gamma \simeq 1.7$), again indicating that the flux recovery is
likely to be associated with the re-emergence of the AGN nuclear
emission. In summary, for this
source there is no doubt that the absorbing and reflecting
regions do not have the same column density and likely belong to different
gaseous structures.

\section{Discussion}

We have presented evidence of the switching--off of the nucleus
in a few Seyfert 2 galaxies based on their changed looks (from
Compton--thin to reflection--dominated or viceversa) when
observed a few years apart. The evidence cannot be considered
conclusive yet, and further investigations, both on the same objects and in 
search of new objects with a similar behaviour, are needed. In the meantime,
let us briefly discuss a few interesting consequences of the proposed
scenario.

\subsection{How many reflection--dominated sources are not
genuine Compton--thick absorbed?}

This is a question that involves, of course, both Seyfert 1s
and Seyfert 2s. It is basically
impossible to estimate the fraction of these transitions 
in obscured AGN, due to the lack of a complete and unbiased
sample of homogeneously defined Seyfert 2 galaxies with sufficient X-ray
temporal and spectroscopic coverage. A XMM--Newton program is ongoing 
to address this question on the complete sample of Compton--thick AGN 
defined in Risaliti et al. (1999). 

\subsubsection{Implications for the optical/X-ray classification
mismatch}

The existence of a population
of Seyfert 1 galaxies with significant X--ray absorption (Maiolino
et al. 2001; Fiore et al. 2001; Della Ceca et al. 2001)
has been recently recognized.
For most of these sources evidence for absorption comes from
the flatness of the X--ray spectrum as derived from a hardness
ratio analysis, rather than from a direct
measurement of the column density, because they are often too faint to
allow for a proper spectral analysis. As the X--ray and optical
observations are usually not simultaneous, it is possible that 
this mismatch is, as least for a fraction of these sources, 
due to a temporary switching--off of the nuclear radiation.
Other explanations are still
possible (e.g. ionization of the X-ray absorbing medium, low gas-to-dust ratio
in the AGN nuclear environment, dust sublimation). However, if the explanation
is indeed in terms of variability, with the sources
caught in different states by the X-ray and optical observations,
one would expect to find also 
sources which were switched--off when observed in the optical and
switched--on when observed in X--rays, namely
X--ray unobscured AGN with a type 2 optical spectrum.
Objects of this kind have indeed been recently discovered
in sparse samples of nearby Seyfert galaxies, as reported by
Pappa et al. (2001) and Panessa et al. (2002). 

\subsubsection{Implications for the cosmic X--ray background}

Another possible implication concerns the modeling of the X--ray Background
(XRB). Popular synthesis models of the XRB (e.g Comastri et al. 1995)
require a significant fraction
of moderately Compton--thick sources, in which the nuclear radiation
can be directly observed at energies of tens of keV. If many of the
reflection--dominated sources will be proven to be simply switched--off
AGN, there may be in principle the need for a revision of the XRB synthesis 
models. As, however, the covering factor of the reflecting matter is by
all evidence pretty large (e.g. Matt et al. 2000), 
it is unlikely that the possible lack of 
intermediate Compton--thick sources will result to be a serious problem.

\subsection{How many cold circumnuclear regions?}

A question that instead directly follows from the observations
of Seyfert 2s which change look, is that of the presence of more
than one cold circumnuclear regions.

For at least two out of four sources in our sample, at least two 
(one thin the other thick) circumnuclear regions are definitely
required. (If they actually correspond to two physically and
geometrically distinct regions or simply to inhomogeneities
in one and the same absorber, it is difficult to say with certainty. 
However, there is
evidence that the Compton--thin absorbers are usually located at much
larger distances than the Compton--thick ones, see below).
For the other two, this is possible as well, but the 
limited bandwidth of the observations when the sources were
reflection--dominated does not permit to rule out Compton--thin
reflection. It is worth noting that in another Compton--thin
AGN, NGC~5506, the presence of a Compton--thick reflector can 
be derived from direct spectral analysis (Matt et al. 2001).
Moreover, other Compton-thick AGN in which the soft X-ray spectrum is 
further absorbed by Compton-thin matter
with $N_H \sim 10^{21-22}$~cm$^{-2}$, are known. The Compton--thin absorber
may be associated with the host galaxy
disk (e.g. Tololo~0109-389; Matt et al. 2003) or with dusty regions 
of enhanced stellar formation (e.g. the  IR ultraluminous galaxy 
NGC6240; Iwasawa \& Comastri 1998; Vignati et al. 1999).

Several authors have already suggested the presence of more than one
cold circumnuclear region in AGN. 
Maiolino \& Rieke (1995), Matt (2000) and Weaver (2001) all
proposed the presence of both a Compton--thick and a Compton--thin
absorber. While all these authors agree that the Compton--thick
absorber should be compact (i.e. the `real' torus) on the base of several
direct and indirect evidence in nearby sources 
(i.e. direct imaging in the Circinus Galaxy,
Sambruna et al. 2001; photoionization models of the reflection spectrum
in Circinus and NGC~1068,
Bianchi et al. 2001, and dynamical mass considerations in the same two
sources, Risaliti et al. 1999), there are several possibilities for
the Compton--thin absorber: the Galactic disc (Maiolino \& Rieke 1995);
dust lanes (Malkan et al. 1998; Matt 2000); starburst clouds (Weaver 
2002). It is well possible that different Compton--thin absorbers 
are present in different sources, or even that 
more than one are simultaneously present in the same source.

\section*{Acknowledgments}

GM and RM acknowledges ASI and MIUR (under
grant {\sc cofin-00-02-36}) for financial support.

{}

\end{document}